\documentclass[pra,11pt,letterpaper,nofootinbib,showpacs,showkeys,eqsecnum
] {revtex4}
\usepackage{amsmath,amssymb,amsfonts,graphicx}

\begin{document}

%\tightenlines

\title{The quadratic speedup in Grover's search algorithm from the entanglement perspective}

\author{Pranaw Rungta}
\email{pranaw@cts.iisc.ernet.in}
 \affiliation{Physics Department, Poornaprajna
Institute of Scientific Research, Devanahalli, Bangalore 562110,
India}

 \altaffiliation{Visiting Scientist: CHEP, Indian Institute
of Science, Bangalore, India. }
\date{\today}
%\maketitle

\begin{abstract}

We analyze the role played by entanglement in the dynamical
evolution of Grover's search algorithm in the space of qubits. We
show that the algorithm can be equivalently described as an
iterative change of the entanglement between the qubits, which
governs the evolution of the initial state towards the target
state, and where the entanglement can be quantified in terms a
single entanglement monotone. We also provide a necessary and
sufficient condition for the quadratic speedup, which illustrates
how the change in the bipartite entanglement of the state after
each iteration determines the corresponding increase in the
probability of finding the target state. This allows us to
reestablish from the entanglement perspective that Grover's search
algorithm is the only optimal pure state search algorithm.
\end{abstract}

\maketitle

\section{Introduction}
\label{sec:intro}

Entanglement---a consequence of the {\it tensor-product} structure
(of subsystems) and superposition in quantum mechanics---has been
slated to be one of the main protagonists in quantum computation:
since an {\it efficient} physical realization of a quantum computer
necessarily requires it to have a {\it scalable} tensor-product
Hilbert space---the space of qubits.

Of course, entanglement is not necessary for quantum computation,
since any computation in the space of qubits can be mapped onto a
unary quantum system (for instance, an atom), where the computation
will be devoid of entanglement---Hilbert spaces of same dimensions
are {\it fungible}~\cite{Blume-Kohout2002}. However, as clearly
illustrated in Ref.~\cite{Blume-Kohout2002}, if a quantum computing
system wants to avoid incurring exponential expense in the form of
some physical resource for the computational problems which require
exponential Hilbert space dimensions for its execution, then it is
mandatory for the system to consist of subsystems (or the degrees of
freedom) such that it's Hilbert space is equivalent to the Hilbert
space of qubits, {\it i.e.}, the number of subsystems must grow
linearly with the number of qubits required in an equivalent quantum
computer. Nonetheless, the requirement of the scalable Hilbert space
is necessary but not sufficient for a scalable quantum computer: it
should further allow efficient implementation of the computational
process itself--- {\it i.e.}, the initialization, control unitary
dynamics, error corrections, and measurements. In a nutshell:
entanglement leads to the saving of exponential {\it spatial}
resources by accessing an arbitrary superposition of $N$ states in
just $n$ ($N=2^n$) subsystems~\cite{Jozsa1998}, hence the motivation
to understand it's {\it temporal} role (number of oracle queries
required) in quantum computation.

Jozsa and Linden~\cite{Jozsa2002} provided a major step towards
understanding the role entanglement plays in speeding the dynamical
evolution of a quantum computational process: a pure state quantum
computation process necessarily requires {\it multipartite
entanglement} which grows with problem size to achieve an
exponential temporal speedup, and if the entanglement is capped to a
fixed number of qubits---independent of the problem size---then the
computation can be classically simulated with an equivalent amount
of classical resources (also see \cite{Vidal2003}); this was
demonstrated for Shor's efficient quantum factoring
algorithm~\cite{Shor1994}.

This paper illustrates the temporal role of entanglement in Grover's
search algorithm~\cite{Grover1997}, and in the process further
explains the result of Jozsa and Linden~\cite{Jozsa2002}. Before we
discuss our results, we first define the problem of searching a
database. The problem is defined in terms of an {\it oracle}: one is
given a very large {\it unstructured} database consisting of $N$
($\gg 1$) items, one has to find a (multiple) marked item(s); the
oracle consists of a function $f(x)$: if one inputs an item $x$ (an
oracle query), it outputs $1$ (yes) if $x$ is a marked item,
otherwise it outputs $0$ (no); to obtain a marked item with the
minimum possible number of queries to the oracle is the search
problem. In the space of qubits, Grover's
algorithm~\cite{Grover1997, Boyer1998} executes the search by using
only $O(\sqrt{N})$ oracle queries, but a classical digital
computer---deterministic or probabilistic---would require $O(N)$
oracle queries on an average. Thus, quantum computers provide a {\it
quadratic} temporal speedup over classical computers even though
they both require $O(\log_{2}N)$ spatial resources to perform the
search. What makes Grover's search algorithm significant is the fact
that it is the only {\it optimal}~\cite{Bennett1997, Zalka1999} pure
state search algorithm.

Grover's search algorithm and the understanding of it which emerges
from our paper can be succinctly described as follows. The initial
state $|S_0\rangle$, an equal superposition of $N$ states
representing an unsorted database with $r$ marked states, is
iteratively rotated towards the target state, where each iteration
consists of the oracle operation followed by an application of the
reflection operation (see Sec.~\ref{sec:algorithm}). Each time the
oracle operation inverts the marked states while leaving the
unmarked states unchanged, it creates a {\it minimal structure} in
the state. This is exploited by the reflection operator by rotating
the state about $|S_0\rangle$ to take one {\it optimal} step closer
to the marked states. The structure is minimal in the sense that the
dynamical evolution of initial state is restricted to the {\it
effective} two dimensional space, {\it i.e.}, the state after the
$k$th iteration is given by
\begin{equation}
|S_k\rangle=A_k|t\rangle+B_k|t_\perp\rangle\;, \label{eq:intro1}
\end{equation}
where $|t\rangle$ represents an equal superposition of all the
marked states, and $|t_\perp\rangle$ represents the same for all the
unmarked states. Thus, each iteration evolves the state simply by
changing $A_k$ and $B_k$, the two real parameters (excluding the
normalization condition).

The corresponding entanglement perspective of the algorithm (see
Sec.~\ref{sec:bipartite}): It is the iterative change of
entanglement between the qubits that drives the initial state
towards the target state. An iterative change arises because the
oracle operation generates entanglement between all the
qubits~\cite{Braunstein2002}, which is then necessarily reduced by
the corresponding reflection operator. The need for the iterative
change in the entanglement can be motivated in the following way.
The database consists of a large number of $n$-qubit states, and a
small subset of which are the desired marked states---thus, each
time the oracle generates the entanglement, it facilitates the
corresponding reflection operator to rotate the resulting state one
step closer to the desired marked states. However, the
consequence---or the limitation---of the dynamical evolution of the
algorithm in the effective two dimensional space translates into the
fact that the entanglement in the algorithm is restricted to a
trivial {\it bipartite} form---{\it i.e.}, it can be quantified by a
{\it single} bipartite measure of entanglement. This is due to the
fact that the Schmidt decompositions of the state after the $k$th
iteration, $|S_k\rangle$, with respect to all the divisions of $n$
qubits into two subsets, will consist of the Schmidt coefficients
which are necessarily functions of $A_k$ and $B_k$ (except for a
constant factor). Now if you fix either $A_k$ or $B_k$, it
simultaneously fixes all the coefficients for all the
decompositions. Moreover, since an entanglement measure is
necessarily defined in terms of the Schmidt coefficients, an
arbitrary Schmidt decomposition and a single measure of entanglement
is thus sufficient to fix the entanglement of the state.

More precisely, it is the change of entanglement in the state
$|S_k\rangle$ due to the $(k+1)$th iteration that determines
${(A_{k+1})^2-(A_k)^2}$, the change in the probability of finding
the marked states from the $k$th to the $k+1$th iteration. This
follows from the equation derived in Sec.~\ref{sec:concurrence2}:
\begin{equation}
C(|S_k\rangle)={1\over 2A_0}{d{A_k^2}\over{d k}}\;, \label{eq:intro}
\end{equation}
where $C(|S_k\rangle)$ represents the {\it
concurrence}~\cite{Rungta2001}---a measure of entanglement---of the
state $|S_k\rangle$. We show that the above equation is a necessary
and sufficient condition for the quadratic speedup, and the
integration of the equation, such that $A_k^2$ changes from $A_0^2$
to $1$, determines the number of the oracle operations required for
the search. This fact allows us to further reestablish from the
entanglement perspective that Grover's search algorithm is the only
optimal pure state search algorithm~\cite{Zalka1999}.

The inference that can be drawn from our paper is that if a
quantum algorithm requires $O(2^{n/M})$ oracle queries for it's
execution, then it optimally exploits $M$ {\it effective}
dimensions of the $2^n$ (the problem size) dimensional Hilbert
space---{\it i.e.}, the pure initial state at all times during the
evolution by the algorithm can be represented in terms of the same
$M$ orthogonal states. This will translate into the optimal
exploitation of $M$-partite entanglement---{\it i.e.}, $M-1$
independent measures of entanglement will govern the evolution of
the algorithm. Therefore, when one oracle query is needed for the
execution of an algorithm (an exponential speedup), just as in
Shor's algorithm~\cite{Shor1994}, then it would necessarily
require optimal utilization of $n$-partite entanglement which
grows with the problem size~\cite{Jozsa2002}. Simply put, the
bipartite entanglement is the complete story of Grover's search
algorithm---the reason for the quadratic versus the desired
exponential speedup---which stems from the inherent inability of
the oracle in generating any global structure. This implies a lack
of multipartite entanglement between the qubits which grows with
problem size, a necessary requirement for the exponential
speedup~\cite{Jozsa2002}. In conclusion: the entanglement allows
{\it simultaneous} saving of spatial and temporal resources when a
quantum algorithm is executed in the space of qubits.

\section{Grover's algorithm: the superposition and interference}
\label{sec:algorithm}

Here we give a brief summary of Grover's algorithm; for further
details see~\cite{Grover1997, Boyer1998, Nielsen2000}. We consider a
database of $N$ ($\gg 1$) elements, and let it contain $r$ ($r\leq
N$) marked elements, which we want to find. The database is mapped
onto the $N$ states of a quantum system:
\begin{equation}
|X_j\rangle\;;\;\; j=1,\ldots N\;. \label{eq:database}
\end{equation}
The first step of the algorithm is to form a equal superposition of
the $N$ states:
\begin{equation}
|S_0\rangle={1\over{\sqrt{N}}}\sum_{j=1}^N|X_j\rangle\;.
\label{eq:supp}
\end{equation}
Let us assume that $r$ is known, and define the target state
$|t\rangle$ as
\begin{equation}
|t\rangle={1\over{\sqrt{r}}}\sum_{j=1}^r|X_j\rangle\;,
\label{eq:target}
\end{equation}
a normalized linear combination of marked states. Similarly, the
nontarget state $|t_\perp\rangle$ represents the same for the
unmarked states:
\begin{equation}
|t_\perp\rangle={1\over{\sqrt{N-r}}}\sum_{j=r+1}^N|X_j\rangle\;.
\label{eq:nontarget}
\end{equation}
By using equations (\ref{eq:target}) and (\ref{eq:nontarget}),
$|S_0\rangle$ can be reexpressed as
\begin{eqnarray}
|S_0\rangle &=& {\sqrt{r\over
N}}|t\rangle+{\sqrt{{N-r}\over N}}|t_{\perp}\rangle \nonumber\\
&\equiv& \sin\theta|t\rangle+\cos\theta |t_{\perp}\rangle\nonumber\\
&\equiv& A_0|t\rangle+ B_0|t_{\perp}\rangle\;. \label{eq:szero}
\end{eqnarray}
The state $|S_0\rangle$ is iteratively evolved to the target state
$|t\rangle$, where each iteration consist of two unitary operations,
the oracle operation $R_O$:
\begin{equation}
R_O=I-2|t\rangle\langle t|\;, \label{eq:oracle}
\end{equation}
followed by the reflection operator $R_{S_0}$:
\begin{equation}
R_{S_0}=2|S_0\rangle\langle S_0|-I\;.
 \label{eq:reflection}
\end{equation}
By definition, the oracle $R_0$~(\ref{eq:oracle}) operation inverts
the marked states, leaving the unmarked states unchanged; and the
reflection operator $R_{S_0}$~(\ref{eq:reflection}) rotates the
state about the hyperplane $|S_0\rangle$, hence the name `{\it
reflection operator}'. The central feature of the algorithm is that
the iterative application of $R_{S_0}R_0$ simply rotates
$|S_0\rangle$ in the effective two dimensional hyper plane,
$\{|t\rangle,|t_\perp\rangle\}$. This can be deduced by their action
as described below. Let the state after $k$th iteration be
\begin{equation}
|S_k\rangle=A_k|t\rangle+ B_k|t_\perp\rangle\;.
\label{eq:kiteration}
\end{equation}
An application of $R_O$ on $|S_k\rangle$ gives
\begin{equation}
|S_k\rangle=-A_k|t\rangle+ B_k|t_\perp\rangle\;, \label{eq:oracleO}
\end{equation}
and a further application of $R_{S_0}$ on the above state gives
\begin{eqnarray}
\label{eq:interference} |S_{k+1}\rangle &=&\Biggl({{N-2r}\over
N}A_k+2{{\sqrt{r(N-r)}}\over N}
B_k\Biggr)|t\rangle+\Biggl({{N-2r}\over N}B_k-2{{\sqrt{r(N-r)}}\over
N}
A_k\Biggr)|t_\perp\rangle\nonumber\\
&\equiv& A_{k+1}|t\rangle+B_{k+1}|t_\perp\rangle\;.
\end{eqnarray}
By solving the recursion relation contained in
Eq.~(\ref{eq:interference}) we obtain~\cite{Boyer1998, Nielsen2000}
\begin{equation}
A_k=\sin(2k+1)\theta\;;\;\;B_k=\cos(2k+1)\theta\;,
\label{eq:recursion}
\end{equation}
which can be verified by mathematical induction. Now, if we choose
the number of iterations $k$ as the nearest integer to
$(\pi/4)\sqrt{N/r}$, then the state $|S_k\rangle\approx|t\rangle$,
and a further measurement in the computational basis will provide
one of the marked states. Moreover, the algorithm can be
appropriately modified such that an arbitrary initial state
$|\psi\rangle$ can be used as the initial state, as long as
$\langle{t}|\psi\rangle\equiv{p}$ is nonzero, to search the database
in $O(\sqrt{1\over p})$ oracle queries~\cite{Grover1998}. The
obvious modification to the standard search algorithm (as described
above) is to replace the reflection operator~(\ref{eq:reflection})
with the following reflection operator:
$R_{|\psi\rangle\langle\psi|}=2|\psi\rangle\langle\psi|-I$.

The search algorithm as presented in this section was independent of
whether the database states $\{|X_j\rangle\}$~(\ref{eq:database})
were of a quantum system, which may or may not be devoid of
entanglement. Moreover, in principle, they could also represent the
states of a classical system which allows superposition, for
example, different modes of a classical electromagnetic wave. We
refer the readers to Ref.~\cite{Lloyd2000} for the implementation of
the oracle and the reflection operations in such systems; but, as
discussed in the introduction, such an implementation necessarily
incurs exponential spatial overhead. For our purpose here, we now
map the algorithm onto the space of qubits.

\subsection{Grover's algorithm in the space of qubits}
\label{sec:subsystems}

The $N$ database elements can be conveniently mapped onto the
$N=2^n$ product states of $n$ qubits:
\begin{equation}
|X_j^n\rangle=|x_1\rangle\otimes\ldots\otimes |x_n\rangle\;;\;\;
j=1,\ldots N\;, \label{eq:qubitdatabase}
\end{equation}
where $|x_i\rangle\in\{|0\rangle,|1\rangle\}$ represents the states
of the $i$th qubits---a two-dimensional Hilbert space ${\cal H}_2$;
and the superscript $n$ denotes that $|X_j^n\rangle\in{\cal
H}_2^{\otimes n}$. The database states
$\{|X_j^n\rangle\}$~(\ref{eq:qubitdatabase}) are known as the {\it
computational basis}. The initial state of the algorithm
$|S_0^n\rangle$~(\ref{eq:supp}) is created by applying Hadamard
transformation $H$~\cite{Nielsen2000} to each of the qubits in the
product state $|0\rangle^{\otimes n}$:
\begin{equation}
|S_0^n\rangle=H^{\otimes n}|0\rangle^{\otimes
n}=\Biggl({{|0\rangle+|1\rangle}\over{\sqrt{2}}}\Biggr)^{\otimes
n}={1\over{\sqrt{N}}}\sum_{j=1}^N|X_j^n\rangle\;. \label{eq:suppq}
\end{equation}
The oracle $R_O$~(\ref{eq:oracle}) is implemented via a conditional
unitary transformation $U_O$ on the computation basis states and an
ancilla qubit $|y\rangle$, which is chosen to be in the state
$(|0\rangle+|1\rangle)/2$:
\begin{equation}
U_O|X_j^n\rangle|y\rangle=|X_j^n\rangle|y\oplus
f(x_j)\rangle={(-1)^{f(x_j)}}|X_j^n\rangle|y\rangle\;,
\label{eq:unot}
\end{equation}
where $\oplus$ denotes addition modulo $2$, and $f(x_j)=1$ if
$|X_j^n\rangle$ is a marked state, otherwise
$f(x_j)=0$~\cite{Nielsen2000}. The action of $U_O$ from the
computational basis perspective---ignoring the ancilla qubit from
the description, as it remains unchanged---reduces to
$R_O=I-2|t^n\rangle\langle t^n|$~(\ref{eq:oracle}). The
implementation of the reflection operator~(\ref{eq:reflection})
follows from the Eq.~(\ref{eq:suppq}):
\begin{equation}
R_{S_0}=H^{\otimes n}(2|0^n\rangle\langle 0^n|-I)H^{\otimes
n}=2|S_0^n\rangle\langle S_0^n|-I\;,
\end{equation}
where $|0^n\rangle\equiv|0\rangle^{\otimes n}$; and by construction
it requires $O(n)$ gates.

The description so far in this section implies that the search
algorithm in the space of qubits requires $O(\sqrt{N/r})$ oracle
queries to perform the search, and it's implementation requires
$O(n)$ physical resources. In any situation where $r$ is not known,
one can either execute the algorithm by averaging over the several
guesses for $r$, and still obtain a marked state in $O(\sqrt{N/r})$
oracle queries; or first estimate $r$ in time $O(\sqrt{N/r})$ by a
quantum algorithm provided in Ref. \cite{Boyer1998}, and then
execute the search.

\section{Entanglement in Grover's algorithm}
\label{sec:bipartite}

Here we quantify the dynamical evolution of the entanglement between
the qubits when the search algorithm is executed in the space of
qubits. We illustrate how the change of entanglement generated after
each iteration plays the central role in the search algorithm. To
facilitate the discussion of the entanglement, and in order to
quantify it, we first introduce a measure called the {\it
concurrence}~\cite{Rungta2001}; the choice to use the concurrence
was dictated by its simplicity (linearity), and in any case, the
essential results would remain the same under an another {\it
entanglement monotone}~\cite{Vidal2000}.

\subsection{Concurrence}
\label{sec:concurrence}

We can always Schmidt decompose an arbitrary bipartite pure state
$|\Psi^{ab}\rangle$ of a ${\cal H}_{D^a}$ and ${\cal H}_{D^b}$
dimensional subsystems:
\begin{equation}
\label{eq:psi} |\Psi^{ab}\rangle=\sum_{j=1}^D
\sqrt{\mu_j}|\Psi_j^a\rangle\otimes|\Psi_j^b\rangle\;,
\end{equation}
where $D\equiv\min(D^a,D^b)$. The squared Schmidt coefficients
$\mu_j$ are the eigenvalues of the reduced density operators,
$\rho^a$ and $\rho^b$, of the two systems, and the vectors
$|\Psi_j^a\rangle$ and $|\Psi_j^b\rangle$ make up the orthonormal
bases that diagonalize the reduced density operators. If all but one
of the Schmidt coefficients are zero, then the state is {\it
separable}, otherwise the state is entangled. The concurrence of the
bipartite pure state $|\Psi^{ab}\rangle$ is simply related to the
{\it purity} of the marginal density operators~\cite{Rungta2001},
\begin{equation}
C(|\Psi^{ab}\rangle)= \sqrt{2(1-{\rm
tr}[(\rho^a)^2])}=\sqrt{2\Biggl(1-\sum_{j=1}^{D}\mu_j^2\Biggr)}=
2\sqrt{\sum_{j<k}\mu_j\mu_k}\;. \label{eq:concurrence}
\end{equation}
$C(|\Psi^{ab}\rangle)$ is conserved under local unitary
transformations because it is a function only of the Schmidt
coefficients.  It varies smoothly from $0$, for pure product states,
to $\sqrt{2(D-1)/D}$, for maximally entangled pure states.

\subsection{The Schmidt decomposition of ${\bf |S_k^n\rangle}$}
\label{subsec:split}

We want to quantify the entanglement of the state generated after
the $k$th iteration of the algorithm. Recall, the state after the
$k$th iteration of the search algorithm is given by
\begin{equation}
|S_k^n\rangle=A_k|t^n\rangle+ B_k|t_\perp^n\rangle\;,
\label{eq:bikiteration}
\end{equation}
where (and henceforth) the superscript $n$ denotes that the states
are $n$-qubit states. We want to Schmidt decompose $|S_k^n\rangle$
with respect to an arbitrary partition of $n$ qubits into two
subsets of (say the first) $l$ qubits and the remaining $(n-l)$
qubits; which will allow us to obtain the concurrence of the state
by the use of Eq.~(\ref{eq:concurrence}).

For the sake of simplicity, we assume that there is a single target
state ($r=1$). We discuss the case of multiple marked states in the
appendix~\ref{sec:general}, where we show that the results for a
single marked state generalizes to the case when there are $r$
marked states. To obtain the Schmidt coefficients corresponding to
the bipartite decomposition of $|S_k^n\rangle$, we define the target
state to be
\begin{equation}
\label{eq:bitarget1} |t^n\rangle = |X_1^l\rangle|X_1^{n-l}\rangle\;,
\end{equation}
where $|X_1^l\rangle$ is a $l$ qubit product
state~(\ref{eq:qubitdatabase}). One can conveniently express the
nontarget state as
\begin{eqnarray}
\label{eq:binontarget} |t^n_\perp\rangle &=& {1\over
{\sqrt{N-1}}}\Biggr\{\sqrt{2^{n-l}-1}|X_1^l\rangle|N^{n-l}\rangle+
\sqrt{2^l-1}|N^l\rangle|X_1^{n-l}\rangle\nonumber\\
&+& \sqrt{(2^l-1)(2^{n-l}-1)}|N^l\rangle|N^{n-l}\rangle\Biggl\}\;,
\end{eqnarray}
where
\begin{eqnarray}
|N^l\rangle     &=& (2^l-1)^{-{1\over2}}\sum_{j=2}^{2^l}
|X_j^l\rangle\label{eq:N1}\\
|N^{n-l}\rangle &=& (2^{n-l}-1)^{-{1\over 2}}\sum_{j=2}^{2^{n-l}}
|X_j^{n-l}\rangle\;.\label{eq:N2}
\end{eqnarray}
Equations (\ref{eq:bitarget1}) and (\ref{eq:binontarget}) imply that
the reduced density matrix $\rho_k^l$ obtained by tracing out
$(n-l)$-qubit states from the density operator
$\rho_k^n\equiv|S_k^n\rangle\langle S_k|$ can be represented in
terms of the two dimensional basis $\{|X_1^l\rangle, |N^l\rangle\}$,
{\it i.e.}, $\rho_k^l$ (or $\rho_k^{n-l}$) will have two non zero
eigenvalues, which are the squared Schmidt coefficients of
$|S_k^n\rangle$; and they are
\begin{equation}
\lambda_{\pm}^l(k)={1\over 2}\pm\Biggl({1\over
4}-\eta^2(A_k-B_k\tan\theta)^2B_k^2\Biggr)^{1\over 2}\;,
\label{eq:Schmidtcoefficients}
\end{equation}
where $\eta$ is a constant that depends on the bipartite
decomposition,
\begin{equation}
\eta=\langle N^{n-l}|\langle
N^l|S_k^n\rangle=\Biggl({(2^l-1)(2^{n-l}-1)\over N-1}\Biggr)^{1\over
2}\;, \label{eq:eta}
\end{equation}
and if $N\gg 1$, then $\eta\approx 1$. Also, one can check that
$\eta$ is proportional to the entanglement of the nontarget state,
{\it i.e.}, $C(|t_\perp^n\rangle)=2\eta$

For the sake of completeness, we provide the Schmidt decomposition
of $|S_k^n\rangle$:
\begin{equation}
|S_k^n\rangle={\sqrt{\lambda_+^l(k)}}|\phi_k^l\rangle_+|\phi_k^{n-l}\rangle_+
+{\sqrt{\lambda_-^l(k)}}|\phi_k^l\rangle_-|\phi_k^{n-l}\rangle_- \;,
\label{eq:schmidrep}
\end{equation}
where
\begin{eqnarray}
|\phi_k^l\rangle_{\pm} &=& {\cal N}\Biggl
\{b|X_1^l\rangle+(\lambda_\pm^l(k)-a)|N^l\rangle\Biggr\}\;,\nonumber\\
|\phi_k^{n-l}\rangle_{\pm} &=& {\cal N}^\prime\Biggl
\{b^\prime|X_1^{n-l}\rangle+(\lambda_\pm^l(k)-a^\prime)|N^{n-l}\rangle\Biggr\}
\;, \label{eq:schmidrep1}
\end{eqnarray}
where ${\cal N}$ and ${\cal N}^\prime$ are the normalization
constants, and
\begin{eqnarray}
a &=& {A_k}^2+{{2^{n-l}-1}\over{N-1}}\;{B_k}^2\;,\nonumber\\
a^\prime &=& {A_k}^2+{{2^l-1}\over{N-1}}\;\;{B_k}^2\;,\nonumber\\
b &=& {\sqrt{2^l-1\over {N-1}}}\;A_k
B_k+{{(2^{n-l}-1)\sqrt{2^l-1}}\over{N-1}}\;{B_k}^2\;,\nonumber\\
b^\prime &=& {\sqrt{2^{n-l}-1\over {N-1}}}\;A_k
B_k+{{(2^l-1)\sqrt{2^{n-l}-1}}\over{N-1}}\;{B_k}^2\;. \label{eq:sbc}
\end{eqnarray}

The above Schmidt representation of $|S_k^n\rangle$ makes an
important point obvious, {\it i.e.}, the main contribution to the
Schmidt coefficients~(\ref{eq:Schmidtcoefficients}), and the
corresponding Schmidt vectors, comes from the term $A_kB_k$. This
arises from the entanglement between the target and nontarget states
generated by the oracle---a fact central for the interpretation of
our results, as discussed in the introduction. Now we use the
Schmidt representation to analyze the dynamical evolution of the
entanglement in the algorithm.

\subsection{The concurrence of ${\bf |S_k^n\rangle}$}
\label{sec:concurrence2}

Equations (\ref{eq:concurrence}) and (\ref{eq:Schmidtcoefficients})
imply that the concurrence of $|S_k^n\rangle$ is given by
\begin{eqnarray}
C(|S_k^n\rangle)
       &=& 2\eta B_k\Biggl( A_k-B_k\tan\theta\Biggr )\label{eq:C1}\\
       &=& 2\eta\sec\theta\sin
       2k\theta\;\cos(2k+1)\theta\label{eq:C2}\\
       &=& \eta{\sec\theta\over 2\theta}\Biggl({\sin
2k\theta\over{\sin(2k+1)\theta}}\Biggr){d{A_k^2}\over{d
k}}\label{eq:C3}\\
       &\approx& \eta{\sec\theta\over 2\theta}{d{A_k^2}\over{d
k}}\label{eq:C4}\\
&\approx& {1\over 2A_0}{d{A_k^2}\over{d k}}\;,\label{eq:C5}
\end{eqnarray}
where the first equality follows from (\ref{eq:concurrence}) and
(\ref{eq:Schmidtcoefficients}); the second inequality is obtained by
using $A_k=\sin(2k+1)\theta$ and
$B_k=\cos(2k+1)\theta$~(\ref{eq:recursion}). Moreover, the second
equality assumes that we are in the first quadrant of the hyperplane
$\{|t\rangle, |t_\perp\rangle\}$ (henceforth our discussion will be
restricted to the first quadrant), which means that all the
quantities in (\ref{eq:C2}) are positive; otherwise one has to put
the absolute value sign in the right hand side of the (\ref{eq:C2}),
since by definition the concurrence~(\ref{eq:concurrence}) is
positive. The third equality~(\ref{eq:C3}) is a different
representation of the second, which follows from
Eq.~(\ref{eq:recursion}); the fourth equality~(\ref{eq:C4}) follows
from fact that ${\sin 2k\theta}/{\sin(2k+1)\theta}\approx 1$, and it
is always positive but less than unity, except for the first few
iterations; the fifth equality~(\ref{eq:C5}) results from the
assumption that $N\gg 1$---the domain where the quadratic speedup is
meaningful---then $A_0\approx\theta $ and $\eta\sec\theta\approx 1$,
in which case the concurrence becomes independent of the bipartite
decomposition.

Equations~(\ref{eq:C1})-(\ref{eq:C5}) quantifies the entanglement
between any $l$ and $n-l$ sets of qubits after the $k$ iteration,
except for a negligible constant factor. The Eq.~(\ref{eq:C5}) is
the main result of the paper: it explains that the search algorithm
exploits the change in entanglement after each iteration to evolve
the initial state to the target state, {\it i.e.}, the change in
$C(|S_k^n\rangle)$ due to the $k+1$th iteration which determines
$(A_{k+1})^2-(A_k)^2$, the change in the probability of finding the
target state from $k$th to $k+1$th iteration. Therefore, it follows
that Eq.~(\ref{eq:C5}) will not only governs the number of oracle
queries needed to search the database, but, as we show in the next
section~\ref{sec:main}, that it is indeed a necessary and sufficient
condition for the quadratic speedup.

The evolution of the concurrence $C(|S_k^n\rangle)$ with the change
in the number of the iterations $k$ has been plotted in Fig~$1$:
here we discuss a some of its salient features of it from the
perspective of the change in the Schmidt coefficients with the
change in $k$, further details can be obtained from
Eqs.~(\ref{eq:Schmidtcoefficients})-(\ref{eq:sbc}). First notice,
Eq.~(\ref{eq:C2}) implies $C(|S_0^n\rangle)=0$ ($k=0$) and
$C(|t^n\rangle)=0$ ($k\approx\sqrt{N}(\pi/4)$), as it should be,
since by choice $|S_0^n\rangle$ and $|t^n\rangle$ are separable,
otherwise $C(|S_k^n\rangle)$ is always nonzero. As $k$ is increased,
the difference between the Schmidt coefficients $\lambda_+^l(k)$ and
$\lambda_-^l(k)$ starts decreasing, therefore $C(|S_k^n\rangle)$
monotonically increases, and attains it's maximum value when
$k\approx{\sqrt N(\pi/8)}$, where
$\lambda_+^l(k)=\lambda_{-}^l(k)=1/{\sqrt 2}$. Further iterations
monotonically decreases $C(|S_k^n\rangle)$, since the difference
between the Schmidt coefficients again starts to increase. When
$C(|S_k^n\rangle)$ approaches zero, it signals that the target state
is being approached, {\it i.e.}, when $\lambda_+^l(k)=0$,
$\lambda_{-}^l(k)=1$ (or $A_k=1$, $B_k=0$).

We now give an alternate description of the change in
$C(|S_k^n\rangle)$ with a change in $k$, which is more relevant for
the purpose of this paper. This description naturally arises from
the entangling properties of the oracle and reflection operator.

\begin{figure}[h]
\begin{center}
\includegraphics[width=8.5cm]{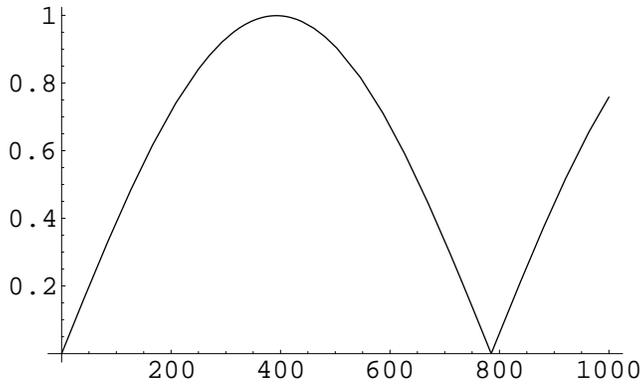}
\end{center}
\vspace{12pt} \caption{Plots of $C(|S_k^n\rangle)$ against $k$ for
$r=100$ and $N=10^8$ ($\eta=1$).} \label{FIG1}
\end{figure}

\subsection{Entangling and Disentangling process in the algorithm}

We first show that the oracle operation of selectively inverting the
target state generates the bipartite entanglement. This can be seen
from the concurrence of the state $R_O|S_k\rangle$:
\begin{eqnarray}
C(R_O|S_k^n\rangle) &=& 2\eta(A_k+B_k\tan\theta)B_k\label{eq:O1}\\
&=& 2\eta\sec\theta\sin2(k+1)\theta\;\cos(2k+1)\theta\label{eq:O2}\\
&=& \eta{\sec\theta\over
2\theta}\Biggl({\sin2(k+1)\theta\over\sin(2k+1)\theta}\Biggr)\;{d{A_k^2}\over{d
k}}\;\label{eq:O3}
\end{eqnarray}
where the second equality~(\ref{eq:O2}) is obtained by substituting
the explicit values of $A_k$ and $B_k$ given in
(\ref{eq:recursion}); the third equality~(\ref{eq:O3}) also follows
from Eq.~(\ref{eq:recursion}), where ${\sin
2(k+1)\theta}/{\sin(2k+1)\theta}\approx 1$, except for the first few
iterations, but because ${\sin 2(k+1)\theta}/{\sin(2k+1)\theta}\geq
{\sin 2k\theta}/{\sin(2k+1)\theta}$, then Eqs.~(\ref{eq:C3}) and
(\ref{eq:O3}) imply that $C_1(R_O|S_k^n\rangle)\geq
C_1(|S_k^n\rangle)$. This can be also seen directly from
\begin{eqnarray}
C_1(R_O|S_k^n\rangle)-C_1(|S_k^n\rangle) &=& 4\eta
B_k^2\tan\theta\\
&=& 4\eta\cos^2(2k+1)\theta\tan^2\theta\;,
\end{eqnarray}
which (in the first quadrant) is always positive but a decreasing
function (see Fig.~$2$).

The corresponding reflection operator increases the probability of
finding the target state, then it should reduce the entanglement
between the qubits. This can be seen by evaluating the difference,
$C_1(|S_{k+1}^n\rangle)-C_1(R_O|S_k^n\rangle)$. The difference is
always negative and decreasing, as shown in Figure $2$ (where we
have plotted the {\it negative} of the difference). Fig. $2$
provides a simple explanation for Fig. $1$: the monotonic increase
of $C(|S_k\rangle)$ as $K$ is increased from $k=0$ to
$k={\sqrt{N/r}}(\pi/8)$ results because, in this range of $k$, each
oracle operation generates more entanglement than the reduction due
to the corresponding reflection operator. Whereas, between
$k={\sqrt{N/r}}(\pi/8)$ and $k\approx{\sqrt{N/r}}(\pi/4)$ the
converse occurs. More importantly, we now show that the iterative
increase and decrease in entanglement due to the oracle and
reflection operator, respectively, is optimal.

\begin{figure}[h]
\begin{center}
\includegraphics[width=8.5cm]{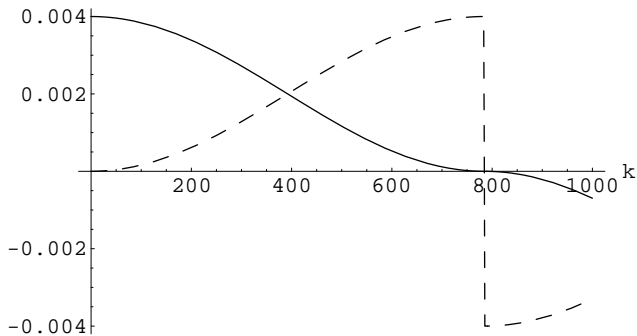}
\end{center}
\vspace{12pt} \caption{Plots of $C(R_0|S_k\rangle)-C(|S_k\rangle)$
(solid line) and $C(R_O|S_k\rangle)-C(R_{S_O}R_O|S_k\rangle)$
(dotted line) against $k$ for $r=100$ and $N=10^8$.} \label{FIG2}
\end{figure}

\section{Quadratic Speedup: The necessary and sufficient~condition}
\label{sec:main}

If an unsorted database of $N$ items has a single marked item, then
Grover's search algorithm finds the marked item in $O(\sqrt{N/r})$
oracle queries. Then, we have already shown, that the state after
the $k$th iteration (or oracle query) $|S_k^n\rangle$ has
entanglement given by Eq.~(\ref{eq:C5}); therefore, it is a
necessary condition to search the database. To show the converse,
assume that there is some (arbitrary) search algorithm which employs
the initial state $|S_0\rangle$~(\ref{eq:suppq}) to search the
database, and it searches in the hyper plane $\{|t\rangle,
|t_\perp\rangle\}$; then our previous discussion implies that the
Schmidt decomposition of the $n$-qubit pure state produced after
each iteration with respect to an arbitrary division of $n$ qubits
into two subsets will involve only two nonzero Schmidt coefficients,
excluding the normalization condition. Let the state generated by
the arbitrary algorithm after the $k$th iteration be
$|S_k^n\rangle^\prime$, and let $\lambda_1^k\equiv\sin^2\phi(k)$ and
$\lambda_2^k=\cos^2\phi(k)$ be its squared Schmidt coefficients.
Now, assume that the concurrence of $|S_k\rangle^\prime$ is given by
$(1/{2A_0})d(A_k)^2/{dk}$, where $A_k^2$ is the probability of
finding the target state in $|S_k^n\rangle^\prime$, and
$A_0=\sin\theta=\sin\phi(k=0)\approx\phi$. This implies
\begin{eqnarray}
{1\over{2\phi}}{d(A_k)^2\over{dk}}&=& C(|S_k^n\rangle^\prime)\nonumber\\
&=& 2\sqrt{\lambda_1^k\lambda_2^k} \nonumber\\
&=&\sin{2\phi(k)}\;, \label{eq:CC}
\end{eqnarray}
where the second equality follows from the definition of the
concurrence~(\ref{eq:concurrence}). Integrating the above equation,
and imposing the initial condition $A_0=\phi$, implies
$A_k^2=\sin^2(2k+1)\phi$, hence the condition
$C(|S_k^n\rangle^\prime)=(1/{2A_0})d(A_k)^2/{dk}$ is necessary and
sufficient to achieve the quadratic speedup.

A natural question presents itself: If an algorithm searches out of
the two dimensional hyperplane, can it do better than than the
quadratic speedup? The search executed out of the hyperplane can
happen if the oracle instead of inverting the target state
introduces a relative phase between the target and nontarget
state---see for example, Grover's fixed point
algorithm~\cite{Grover2005,Tulsi2006}. Then it is easy to show that
it will imply $C(|S_k\rangle)<(1/{2\theta})d(A_k)^2/{dk}$, because
the projection in the hyper plane will always reduce the
concurrence~\cite{Rungta2007}, therefore it will provide no temporal
advantage over Grover's search algorithm. Moreover, one can apply an
operation which is a more general operation than the reflection
operation, {\it i.e.}, it is not restricted to the effective two
dimensions of the search algorithm. Then, as shown explicitly by
Zalka~\cite{Zalka1999}, nothing additional can be gained: the
shortest distance between the target and nontarget state,
dynamically speaking, lies in the two dimensional hyperplane
$\{|t\rangle, |t_\perp\rangle\}$, {\it i.e.}, along the
geodesic---hence the reflection operator is optimal. This can also
be argued from the entangled perspective as follows: although the
entanglement of the state which rotates in a multidimensional
hyperplane would involve more than one measure of entanglement---it
won't be fixed just by the concurrence---but it is the oracle
operation that guides the evolution to the desired target state, and
by definition, it is optimally restricted in the two dimensional
hyperplane.

To show that, as far as the pure state search is concerned, Grover's
search algorithm is the only optimal search algorithm, we have to
show that the quadratic speedup is also {\it asymptotically}
optimal~\cite{Bennett1997}, {\it i.e.}, for any number of oracle
queries, and for arbitrary reflection operators.

\subsection{Asymptotic optimality}

Our proof of the {\it asymptotic} optimality is analogous to the
proof in Ref.~\cite{Bennett1997}. Suppose we want to search an
unsorted database of $N$ items with a single marked item. We map the
problem onto a $N$ dimensional space of $n$ ($2^n=N$) qubits.
Consider an arbitrary initial $n$-qubit pure state $|\psi^n\rangle$,
which contains a single marked state $|t^n\rangle$. $|\psi^n\rangle$
is evolved by invoking the oracle, $R_O^t=2|t\rangle\langle t|-I$,
as follows:
\begin{equation}
|\phi_T^n\rangle_t=U_T R_O^tU_{T-1}R_O^t\ldots
U_1R_O^t|\psi^n\rangle\;.
\end{equation}
where the unitary operators $U_j$'s are arbitrary, and the
probability of finding the target state after $T$ oracle queries is
given by ${A_T^t}^2=|\langle t^n|\phi_T^n\rangle_t|^2$. The idea of
the proof is to compare the above evolution to the case when
$|\psi^n\rangle$ is evolved without invoking the oracle $R_O^t$:
\begin{equation}
|\phi_T^n\rangle=U_T\ldots U_1|\psi^n\rangle\;,
\end{equation}
in which case, let the probability of finding the marked state after
$T$ oracle queries is given by ${A_T}^2=|\langle
t^n|\phi_T^n\rangle|^2$. The main part of the proof works by
obtaining an upper bound for the difference $|{A_T^t}^2-{A_T}^2|$,
and averaging over $N$ linearly independent choices of the marked
states, since one can always design a special algorithm which is
suited for a particular choice of the marked state. The upper bound
can be obtained by considering Eq.~(\ref{eq:C5}):
\begin{eqnarray}
{1\over{A_0^2}}\Biggl({d\over{dk}}\sum_t|{A_k^t}^2-{A_k}^2|\Biggr )
&\leq&
\sum_t|C(|\phi_k^n\rangle_t)-C(|\phi_k^n\rangle)|\\
&\leq& \sqrt{2}\;, \label{eq:difference}
\end{eqnarray}
where ${A_0}^2$ is the probability of finding the marked state in
the initial state $|\psi^n\rangle$, and the second inequality is
obtained from the fact that the concurrence is bounded by
$\sqrt{2(N-1)/N}\approx \sqrt{2}$ (see Sec.~\ref{sec:concurrence}).

Integration of the above equation~(\ref{eq:difference}) provides the
upper bound:
\begin{equation}
\sum_t|{A_k^t}^2-{A_k}^2|\leq\sqrt{2}\;T \sqrt{N}\;, \label{eq:o1}
\end{equation}
where we have used $A_0=\sqrt{1/N}$. Now to complete the proof one
obtains the lower bound by considering a worst case scenario: we
want to invoke the oracle large enough times such that we should be
able to distinguish sufficiently  via a measurement all
$|\phi_k^n\rangle_t$, {\it i.e.}, for $N$ linearly independent
choices of $|t^n\rangle$, which in turn implies that for some large
$k$, and for some fixed $\epsilon$, ${A_k^t}^2-{A_k}^2\geq\epsilon$;
moreover, this should be true for all $N$ choices of the marked
states:
\begin{equation}
\sum_{t=1}^N|{A_k^t}^2-{A_k}^2|\geq \epsilon N\;. \label{eq:o2}
\end{equation}
Then, (\ref{eq:o1}) and (\ref{eq:o2}) implies $T\geq
 O(\sqrt{N})$.

Therefore, quadratic speedup is also asymptotically
optimal~\cite{Bennett1997}. Thus, Grover search algorithm is the
only optimal algorithm for searching an unsorted database with a
pure state. The question: Can parallel quantum computation improve
the quadratic speedup?

\section{Entangled parallel quantum computing}

Here we show that when Grover's search algorithm is executed by $l$
entangled computers---or $l$ different sets of $n$ qubits---then it
provides a certain linear advantage, {\it i.e.}, they can produce
$l$ copies of the target state, where each copy has $r$ marked
states, with just $O(\sqrt{N/r})$ oracle queries. The advantage can
be motivated in the following way. The result of the search
algorithm is an equal superposition of all the marked states, and a
further measurement in the computational basis provides one of the
marked states. Suppose each of $r$ marked states encodes different
information, therefore, to extract all the information, the search
algorithm needs to be executed (say) $l$ times to produce $l$ copies
of the target state, which will require $O(l\sqrt{N/r})$ oracle
queries.

We now show that the search with a multipartite entangled state can
provide the linear advantage. We denote the computational basis
states of the $k$th quantum computer as $\{|X_j^n\rangle_k\}$,
$j=1\ldots N$, and the initial state of the $l$ quantum computers is
a generalized GHZ state:
\begin{equation}
|S_0^n\rangle_l={1\over {\sqrt
N}}\sum_j^N|X_j^n\rangle_1\otimes\ldots|X_j^n\rangle_l\;,
\end{equation}
Now define the target state of $l$ computers as
\begin{equation}
|t^n\rangle_l={1\over{\sqrt{r}}}\sum_{j=1}^r|X_j^n\rangle_1\otimes\ldots\otimes|X_j^n\rangle_l\;,
\label{eq:mtarget}
\end{equation}
and similarly, the nontarget state $|t_\perp^n\rangle_l$ can be
defined:
\begin{equation}
|t_\perp^n\rangle_l={1\over{\sqrt{N-r}}}\sum_{j=r+1}^N|X_j^n\rangle_1\otimes\ldots\otimes|X_j^n\rangle_l\;.
\label{eq:mnontarget}
\end{equation}
By using equations (\ref{eq:mtarget}) and (\ref{eq:mnontarget}),
$|S_0^n\rangle_l$ can be rewritten, just as in the standard search
algorithm (see Sec.~\ref{sec:algorithm}):
\begin{eqnarray}
|S_0^n\rangle_l &=& {\sqrt{r\over
N}}|t^n\rangle_l+{\sqrt{{N-r}\over N}}|t_{\perp}^n\rangle_l \nonumber\\
&\equiv& \sin\theta|t\rangle_l+\cos\theta_l |t_{\perp}\rangle\nonumber\\
&\equiv& A_0|t\rangle_l+ B_0|t_{\perp}\rangle_l\;. \label{eq:mszero}
\end{eqnarray}
Now, the standard search algorithm is executed by a single computer
(say, the first), and the rest of of computers do nothing, {\it
i.e.}, the oracle operation $R_O^l$ for the $l$ entangled computers
is made of the product of $l$ unitary operations, where the first
unitary operation is the standard oracle
operation~(\ref{eq:oracle}), and the rest of the unitaries are the
respective identity operators. Similarly, the reflection operator
$R_{S_O}^l$ for the computers can be defined. It is straight forward
to show that after $m=(\pi/4)\sqrt{N/r}$ iterations the {\it local}
density operators of all the computers reduces to
\begin{equation}
{1\over r}\sum_{j=1}^r|X_j^n\rangle\langle X_j^n|\;.
\end{equation}
Notice, that the linear advantage is achieved via the entanglement
in the initial state, although the joint unitary operations are
local.

\section{Conclusion}

Our discussion in this paper was restricted to the search algorithm,
where the initial pure state was an equal superposition of all
database states; however, it can be easily generalized to an
arbitrary pure initial state. Moreover, if the search is performed
with a mixed initial state, then our result, Eq.~(\ref{eq:C5}), has
to be optimized not only over all the ensemble decomposition of the
state, but also over all reflection operators. This, we will
consider in a future publication.

Since the entanglement in the search algorithm is of the limited
bipartite form, it explains why the quadratic (temporal) speedup
can be achieved if the algorithm is implemented in a classical
system which allows superposition, as explicitly shown by
Lloyd~\cite{Lloyd2000}. This is further reflected by the fact that
even a classical digital computer can trivially simulate the
algorithm. However, as the number of bipartite entanglement
measures required for a quantum computational process grows (the
multipartite-ness of entanglement increases), then the temporal
expense incurred by the simulation of the computational process by
a classical system, irrespective of whether the superposition is
allowed or not, will necessarily grow; more importantly, the
simulation will have no efficient {\it realistic}
description~\cite{Blume-Kohout2002} of the computational process
in the limit where the multipartite entanglement needed for the
computations grows with the problem size~\cite{Jozsa2002}. After
all, what distinguishes quantum computation---whether there is
entanglement or not---from classical computation---where
superposition is allowed or not---is that an arbitrary quantum
state and it's dynamics has no efficient {\it realistic}
description: the reason, perhaps, that makes quantum computation
inherently more powerful than classical
computation~\cite{Blume-Kohout2002}.

The main contribution of this paper is an explicit illustration of
the dynamical role of entanglement in the search algorithm: the
change of entanglement after each iteration governs the evolution of
the initial state. Although the entanglement in the algorithm is
limited, it is optimally exploited by the search algorithm. It is
precisely due to the limited bipartite entanglement in the
algorithm, that the algorithm leads to the quadratic speedup. The
bipartite entanglement itself is a consequence of the oracle: which
by definition, optimally restricts the evolution of the algorithm to
the two effective Hilbert space dimensions. The simplicity of the
algorithm does succeed in illuminating that entanglement---perhaps,
entangling operations in the case of a mixed state
computation---plays an essential role in saving both the spatial and
temporal resources when exponential Hilbert space dimensions are
required for the execution of a quantum computational process. Thus,
entanglement is indispensable for a scalable quantum computer.

\appendix

\label{sec:general}

\section{Multiple marked states}

Here we show that all our previous analysis of the role of
entanglement in Grover's search algorithm, where we had assumed that
there is a single marked state, can be generalized when there are
$r$ arbitrary marked states, {\it i.e.}, the target state is given
\begin{equation}
|t^n\rangle={1\over{\sqrt{r}}}\sum_{j=1}^r|X_j^n\rangle\;,
\label{eq:target2}
\end{equation}
where $|X_j^n\rangle$ are computational basis
states~(\ref{eq:qubitdatabase}). This implies that $|S_k^n\rangle$
may not be separable in the arbitrary bipartite decomposition of the
$n$ qubits. Recall, the state after the $k$th iteration is given by
\begin{equation}
|S_k^n\rangle=A_k|t^n\rangle+ B_k|t_\perp^n\rangle\;,
\label{eq:bikiterationg}
\end{equation}
and we want to obtain it's concurrence corresponding to the
bipartite division of $n$ qubits into $l$ qubits and $n-l$ qubits.
Let's assume, without any loss of generality, that the $r$ marked
states belong in the set of database states:
\begin{equation}
M\equiv
\Bigl\{|X_j^l\rangle|X_k^{n-l}\rangle\Bigr\}\;;\;j=1,\ldots,p\;,\;k=1,\ldots,q\;,
\label{eq:markedspace}
\end{equation}
where $p\leq r$, $q\leq r$, and $pq\geq r$; and $|X_j^l\rangle$ and
$|X_j^{n-l}\rangle$ are $l$-qubit and $(n-l)$-qubit computational
basis states, respectively. To obtain the concurrence of
$|S_k^n\rangle$, we first define a convenient basis:
\begin{eqnarray}
|T^l\rangle &=& {1\over \sqrt{p}}\sum_{j=1}^p|X_j^l\rangle\;,\label{eq:b1}\\
|T^{n-l}\rangle &=& {1\over
\sqrt{q}}\sum_{j=1}^q|X_j^{n-l}\rangle\;,\label{eq:b2}\\
|N^l\rangle &=& {1\over
\sqrt{2^l-p}}\sum_{j=1}^{2^l-p}|X_j^l\rangle\;,\label{eq:b3}\\
|N^{n-l}\rangle &=& {1\over
\sqrt{2^{n-l}-q}}\sum_{j=1}^{2^{n-l}-q}|X_j^{n-l}\rangle\;,\label{eq:b4}
\end{eqnarray}
where $|T^l\rangle$ and $|T^{n-l}\rangle$ are a sum of all
$|X_j^l\rangle$'s and $|X_j^{n-l}\rangle$'s, respectively, which are
contained in $M$; similarly, $|N^l\rangle$ and $|N^{n-l}\rangle$ are
sum of all $|X_j^l\rangle$'s and $|X_j^{n-l}\rangle$'s,
respectively, that are not in $M$.

Notice, we can always express $|t^n\rangle$ as follows:
\begin{equation}
|t^n\rangle=\sum_{j=1}^q{\sqrt{p_j\over
r}}|P_j^l\rangle|X_j^{n-l}\rangle\;,
\end{equation}
where $\sum_jp_j=r$, and
\begin{equation}
|P_j^l\rangle={1\over {\sqrt p_j}}\sum_{j=1}^{p_j}|X_j^l\rangle\;,
\label{eq:targetg}
\end{equation}
where the sum is over $p_j$ number of states $|X_j^l\rangle$ which
belongs in $M$, and $\langle P_j^l|P_k^l\rangle\equiv
p_{jk}/\sqrt{p_jp_k}$. The nontarget state $|t_\perp^n\rangle$, by
the use of Eqs.~(\ref{eq:b1})-(\ref{eq:b4}), can be conveniently
expressed as
\begin{eqnarray}
|t_\perp^n\rangle &=& {1\over \sqrt{N-r}}\Biggl(
{\sqrt{pq}}|T^l\rangle|T^{n-l}\rangle-{\sqrt{r}}|t^n\rangle\nonumber\\
&+& \sqrt{q(2^l-p)}|N^l\rangle|T^{n-l}\rangle +
{\sqrt{(2^l-p)(2^{n-l}-q)}}|N^l\rangle|N^{n-l}\rangle\Biggr)\;.
\label{eq:nontargetg}
\end{eqnarray}
By substituting (\ref{eq:targetg}) and (\ref{eq:nontargetg}) in
(\ref{eq:bikiterationg}), one obtains
\begin{eqnarray}
|S_k^n\rangle &=& \Bigl(A_k-B_k\tan\theta\Bigr)\sum_{j=1}^q
{\sqrt{p_j\over
r}}|P_j^l\rangle|X_j^{n-l}\rangle\nonumber\\
&+& {B_k\over\sqrt{N-r}}
\Biggl({\sqrt{pq}}|T^l\rangle|T^{n-l}\rangle +
\sqrt{q(2^l-p)}|N^l\rangle|T^{n-l}\rangle\nonumber\\
&+& {\sqrt{(2^l-p)(2^{n-l}-q)}}|N^l\rangle|N^{n-l}\rangle\Biggr)\;,
\end{eqnarray}
where we have substituted $\tan\theta=\sqrt{r/{(N-r)}}$. By tracing
out the $(n-l)$-qubit states from the density operator
$\rho_k^n=|S_k^n\rangle\langle S_k^n|$, one can obtain $\rho_k^l$;
which can then be substituted in the expression
\begin{equation}
C(|S_k^n\rangle)\equiv\sqrt{2[1-{\rm
tr}\Bigl((\rho_k^l)^2\Bigr)]}\;,
\end{equation}
to show that the square concurrence of $|S_k^n\rangle$ is given by
\begin{eqnarray}
C^2(|S_k^n\rangle) &=&{4(2^l-p)(2^{n-l}-q)\over
{N-r}}(A_k-B_k\tan\theta)^2B_k^2\\\label{eq:c1} &+&
(A_k^2-B_k^2\tan\theta)^2C^2(|t^n\rangle)\;, \label{eq:squareC}
\end{eqnarray}
where the first term represents the entanglement generated by the
search algorithm after the $k$the iteration, but the second term is
simply a byproduct of the initial entanglement, since it is not
affected by the oracle operation. Therefore, when analyzing the role
of entanglement in the search algorithm, it is necessary to ignore
the second term in the above equation. If we do so, then the
concurrence of the state $|S_k^n\rangle$ generated by the algorithm
is given by
\begin{equation}
C(|S_k^n\rangle)=2\eta^\prime(A_k-B_k\tan\theta)B_k\;,
\end{equation}
where
\begin{equation}
\eta^\prime=\langle N^{n-l}|\langle
N^l|S_k^n\rangle=\Biggl({(2^l-p)(2^{n-l}-q)\over N-r}\Biggr)^{1\over
2}\;, \label{eq:eta}
\end{equation}
and when there is a single target state, {\it i.e.}, when
$C(|t^n\rangle)=0$, $p=q=1$, and $\eta^\prime\rightarrow\eta$, it
reduces (so does Eq.~(\ref{eq:squareC}) to Eq.~(\ref{eq:C1}). Thus,
the dynamical evolution of the algorithm, when there are $r$ marked
states, in terms of the concurrence can be similarly expressed as
\begin{eqnarray}
C(|S_k^n\rangle)
       &=& \eta^\prime{\sec\theta\over 2\theta}\Biggl({\sin
2k\theta\over{\sin(2k+1)\theta}}\Biggr){d{A_k^2}\over{d
k}}\label{eq:C3}\\
&\approx& {1\over 2A_0}{d{A_k^2}\over{d k}}\;, \label{eq:CM}
\end{eqnarray}
where $A_0=\sin\theta=\sqrt{r/N}$. Below, we illustrate the
derivation of Eq.~(\ref{eq:squareC}) via a simple example, where
there is a little loss of generality, and it is more instructive.

\subsection{An entangled target state}

Suppose we Schmidt decompose $|S_k^n\rangle$ with respect to the
division of $n$ qubits into $n-1$ qubits and the remaining $n$th
qubit. There is no loss of generality here, because it is sufficient
to quantify the entanglement of $|S_k^n\rangle$, otherwise it would
mean that more than two parameters $A_k$ and $B_k$ (excluding the
normalization condition) is being changed after each iteration.
Moreover, the Schmidt decomposition of $|S_k^n\rangle$ with respect
to $(n-1)$-qubit $n$th-qubit states can be used to prove that
Eq.~(\ref{eq:CM}) is necessary and sufficient for the quadratic
speedup when there are multiple marked states, just as we did for a
single marked state (see Sec.~\ref{sec:main}).

Now, suppose that the target state $|t^n\rangle$ in the bipartite
decomposition is given by
\begin{equation}
|t^n\rangle={1\over{\sqrt{r}}}\Bigl(\sqrt{p} |P^{n-1}\rangle\otimes
|0\rangle +\sqrt{q}|Q^{n-1}\rangle\otimes|1\rangle \Bigr)\;,
\label{eq:e1}
\end{equation}
where
\begin{equation}
|P^{n-1}\rangle={1\over{\sqrt{p}}}\sum_{i=1}^p |X_i^{n-1}\rangle\;,
\end{equation}
\begin{equation}
|Q^{n-1}\rangle={1\over{\sqrt{q}}}\sum_{p+1}^r |X_i^{n-1}\rangle\;,
\end{equation}
where $p+q=r$. We further assume for the sake of simplicity that
$|P^{n-1}\rangle$ and $|Q^{n-1}\rangle$ are orthogonal, {\it i.e.},
Eq.~(\ref{eq:e1}) is the Schmidt decomposition of $|t^n\rangle$. If
$|P^{n-1}\rangle$ and $|Q^{n-1}\rangle$ are not orthogonal, which
will generally be the case in the computation basis, then one can
start with the decomposition of $|t^n\rangle$ as given in
Eq.~(\ref{eq:targetg}). $|t^n\rangle$~(\ref{eq:e1}) can be
conveniently reexpressed as
\begin{equation}
|t^n\rangle= \sin\phi\; |P^{n-1}\rangle\otimes |0\rangle +
\cos\phi\;|Q^{n-1}\rangle\otimes|1\rangle \;, \label{eq:target2}
\end{equation}
where $\sin\phi=\sqrt{p/r}$ and $\cos\phi=\sqrt{q/r}$ are the
Schmidt coefficients. If either $p$ or $q$ is zero (or $\phi=0$),
then the target state is separable, otherwise it is entangled. The
entanglement in $|t^n\rangle$ is
\begin{equation}
C(|t^n\rangle)=\sin2\phi\;,
\end{equation}
obtained by using Eq.~(\ref{eq:concurrence}). This implies that
$|t_\perp^n\rangle$ can be decomposed as
\begin{eqnarray}
|t_\perp^n\rangle & = & {1\over{\sqrt{N-r}}}\Biggl(\sqrt{p}\;
|P^{n-1}\rangle|1\rangle
+\sqrt{q}\;|Q^{n-1}\rangle|0\rangle+\sqrt{{N/2}-r}\;|N^{n-1}\rangle(|0\rangle+|1\rangle)
\Biggr)\nonumber\\
&= & \tan\theta\Biggl( \sin{\phi} |P^{n-1}\rangle|1\rangle +
\cos{\phi}|Q^{n-1}\rangle|0\rangle\Biggr) +
{\sqrt{1-\tan^2{\theta}}}\Biggl(
|N^{n-1}\rangle({|0\rangle+|1\rangle\over\sqrt{2}})\Biggr)\;,
\label{eq:nontarget2}
\end{eqnarray}
where $|N^{n-1}\rangle$ is defined to be
\begin{equation}
|N^{n-1}\rangle={1\over{\sqrt{{2^{n-1}}-r}}}\sum_{r+1}^{N/2}
|X_i^{n-1}\rangle\;,
\end{equation}
therefore, by construction, $|N^{n-1}\rangle$ is orthogonal to
$|P^{n-1}\rangle$ and $|Q^{n-1}\rangle$. The concurrence of the
nontarget state is given by
\begin{equation}
C(|t_\perp^n\rangle)=\tan{\theta}\sqrt{2-\tan^2\theta(1+\cos^2{2\phi})}\;.
\end{equation}
The initial state $|S_0^n\rangle$~(\ref{eq:suppq}) is by definition
separable in all the bipartite decomposition:
\begin{equation}
\label{eq:szerobipartite}
|S_0^n\rangle=\Biggl(\sqrt{2}\sin\theta\sin\phi|P^{n-l}\rangle+\sqrt{2}\sin\theta\cos\phi|Q^{n-l}\rangle
+{\sqrt{\cos2\theta}}|N^{n-l}\rangle\Biggr)\Biggl({|0\rangle+|1\rangle\over\sqrt
{2}}\Biggr)\;,
\end{equation}
which is obtained by substituting (\ref{eq:target2}) and
(\ref{eq:nontarget2}) in Eq.~(\ref{eq:bikiterationg}).

The state after $k$th iteration, by using equations
(\ref{eq:target2}) and (\ref{eq:nontarget2}), can be expressed as
\begin{eqnarray}
|S_k^n\rangle &=&
A_k\biggl(\sin\phi|P^{n-l}\rangle|0\rangle+\cos\phi|Q^{n-l}\rangle|1\rangle\biggr)\nonumber\\
&+&
B_k\tan\theta\Biggl(\sin\phi|P^{n-l}\rangle|1\rangle+\cos\phi|Q^{n-l}\rangle|0\rangle\Biggr)\nonumber\\
&+&
B_k\sqrt{1-{\tan^2}\theta}\Biggl(|U^{n-l}\rangle{(|0\rangle+|1\rangle)\over
 2}\Biggr)\;.
\label{eq:sk}
\end{eqnarray}
The two eigenvalues of the reduced density operator of the $nth$
qubit, obtained by tracing out the states $|P^{n-l}\rangle$,
$|Q^{n-l}\rangle$ and $|U^{n-l}\rangle$, can be used in
Eq.~(\ref{eq:concurrence}) to show that the square concurrence of
$|S_k^n\rangle$ given by
\begin{equation}
C_k^2(|S_k^n\rangle)= C_1^2(|S_k^n\rangle)+C_2^2(|S_k^n\rangle)\;,
\end{equation}
where
\begin{eqnarray}
C_1(|S_k^n\rangle) &=&
2(A_k-B_k\tan\theta)B_k\sqrt{(1-\tan^2\theta)}\\\label{eq:c1} &=&
2(A_k-B_k\tan\theta)B_k\sqrt{2({2^{n-1}}-r)\over{N-r}}\;,
\label{eq:exc1}
\end{eqnarray}
and
\begin{eqnarray}
C_2(|S_k^n\rangle) &=&
(A_k^2-B_k^2\tan\theta)\sin2\phi\\
&=& (A_k^2-B_k^2\tan\theta)C(|t^n\rangle)\;.
\end{eqnarray}
Notice, $C_1(|S_k\rangle)$ and $C_2(|S_k\rangle)$ has the same form
as in Eq.~(\ref{eq:squareC}). $C_1(|S_k\rangle)$ is term which is
the concurrence generated by the search algorithm, and has been
discussed in Sec.~\ref{sec:concurrence}. We restrict our attention
to the properties of $C_2(|S_k^n\rangle)$; it is zero for the
initial state (the initial state is separable by definition),
otherwise nonzero everywhere else; it monotonically increases and
attains its maximum value of $\sin2\phi$---the initial concurrence
of the target state. Thus, $C_2(|S_k^n\rangle)$ is simply a
byproduct of the initial entanglement of the target state. The above
example give us the opportunity to illustrate the special, but
interesting, case $r=N/4$~(\cite{Boyer1998}). It highlights in a
rather more dramatic fashion the entangling and disentangling nature
of the oracle and the refection operator, respectively.

\subsubsection{r={N/4}}

When $r={N/4}$, then it requires just one oracle query to search the
database---classically, it would require two oracle queries. If
$r={N/4}$, then Eq.~(\ref{eq:szero}) implies $A_0=1/2$ and
$B_0={{\sqrt 3}/2}$, thus the initial state expressed in the
bipartite form in Eq.~(\ref{eq:szerobipartite}) reduces to
\begin{equation}
|S_0\rangle={1\over\sqrt{2}}\Biggl(\sin\phi|P\rangle+\cos\phi|Q\rangle
+|U\rangle\Biggr)\Biggl({|0\rangle+|1\rangle\over\sqrt {2}}\Biggr)
\end{equation}
Then the action of $R_O$ on the above state generates a maximally
bipartite entangled $n-1$-qubit and $n$th qubit state:
\begin{equation}
R_O|S_0^n\rangle={1\over\sqrt{2}}\Biggl(\sin\phi|P^{n-l}\rangle-\cos\phi|Q^{n-l}\rangle\Biggr)
\Biggl({|0\rangle-|1\rangle\over\sqrt
{2}}\Biggr)\;+\;{1\over\sqrt{2}}|U^{n-l}\rangle\Biggl({|0\rangle+|1\rangle\over\sqrt
{2}}\Biggr)\;,
\end{equation}
{\it i.e.}, $C_1(R_O|S_0^n\rangle)=1$. Again notice, it is the the
oracle which creates entanglement between the target and nontarget
states. Now the action of $R_{S_0}$ on the above state reduces it to
the target state:
\begin{equation}
R_{S_0}R_O|S_0^n\rangle=|t^n\rangle\;,
\end{equation}
then $C_1(R_{S_0}R_O|S_0^n\rangle)=0$; thus, the reflection operator
reduces entanglement in the search algorithm.

\section*{Acknowledgements}

The author thanks Apoorva D Patel for useful discussions on
Grover's search algorithm, and quantum computation in general.

\end{document}